\documentclass[amssymb,aps,preprint]{revtex4}
\usepackage{graphicx}

\begin{document}
\bibliographystyle{prsty}
\title{Quasiparticle dynamics in overdoped Bi$_{1.4}$Pb$_{0.7}$Sr$_{1.9}$CaCu$_{2}$O$_{8+\delta}$:
Coexistence of superconducting gap and pseudogap below
$T_{c}$}
\author{Saritha K. Nair}
\affiliation{Division of Physics and Applied Physics, School of
Physical and Mathematical Sciences, Nanyang Technological
University, 637371 Singapore}
\author{X. Q. Zou}
\affiliation{Division of Physics and Applied Physics, School of
Physical and Mathematical Sciences, Nanyang Technological
University, 637371 Singapore}
\author{J.-X. Zhu}
\affiliation{Los Alamos National Laboratory, Los Alamos, New
Mexico 87545, USA}
\author{C. Panagopoulos}
\affiliation{Division of Physics and Applied Physics, School of
Physical and Mathematical Sciences, Nanyang Technological
University, 637371 Singapore} \affiliation{Department of
Physics, University of Crete and FORTH, 71003 Heraklion,
Greece}
\author{S. Ishida}
\affiliation{Department of Physics, The University of Tokyo,
Bunkyo-ku, Tokyo 113-0033, Japan}
\author{S. Uchida}
\affiliation{Department of Physics, The University of Tokyo,
Bunkyo-ku, Tokyo 113-0033, Japan}
\author{Elbert E. M. Chia}
\affiliation{Division of Physics and Applied Physics, School of
Physical and Mathematical Sciences, Nanyang Technological
University, 637371 Singapore}
\date{\today}

\begin{abstract}
Photoexcited quasiparticle relaxation dynamics in overdoped
Bi$_{2}$Sr$_{2}$CaCu$_{2}$O$_{8+\delta}$ ($T_{c}$=65~K, hole
doping $p$=0.22) single crystal is investigated as a function
of temperature. We provide evidence of a $\sim$22~meV pseudogap
($T^{\ast}$$\approx$100~K) at this doping level. Our data
support the scenario where both the superconducting gap and
pseudogap coexist in the superconducting state. Our results
also suggest an increased scattering rate between electrons and
spin fluctuations as the sample enters the pseudogap phase.
\end{abstract}

\maketitle All hole-doped cuprate high-temperature
superconductors (HTSCs) exhibit an unusual normal state that is
characterized by the opening of a gap in the electronic
spectrum, at a temperature $T^{\ast}$ above the superconducting
(SC) transition temperature $T_{c}$. Much theoretical and
experimental effort has been spent in ascertaining the origin
of this gap, called the pseudogap (PG) \cite{Timusk1999}, for
the answer may prove crucial in the understanding of
high-$T_{c}$ superconductivity. A fundamental issue regarding
the PG phase is \cite{Fischer2007}: does it compete with, is
unrelated to, or is a precursor of, superconductivity? Related
to this is the number of energy gaps below $T_{c}$: a single
energy gap would imply that the PG is a precursor state, while
two gaps would suggest that the PG is a competing or coexisting
phase \cite{Panagopoulos98}. Another issue is the understanding
of the HTSC phase diagram: does the $T^{\ast}$ line merge with
$T_{c}$ on the overdoped side, or does it cross the SC dome and
falls to zero at a quantum critical point? In the former case,
does the PG phase coexist with the SC phase below $T_{c}$, or
is it a precursor to superconductivity by smoothly evolving
into the SC phase below $T_{c}$?

A variety of experimental techniques have sought to answer some
of these questions. The existence of the PG in underdoped
hole-doped HTSCs is now not in doubt, but the issue is not so
clear in the overdoped regime. Angle-resolved photoemission
(ARPES) data in Bi-2212 \cite{Damascelli2003} revealed a
``peak-dip-hump" feature in the SC phase which persists above
$T_{c}$ in the PG phase. Two energy scales were associated with
the PG --- a low-energy one given by the location of the
leading-edge midpoint (before the ``peak"), and a high-energy
one given by the position of a broad peak (``hump") near the
($\pi$,0) point. The low-energy PG smoothly evolves into the SC
gap upon going from the underdoped to the overdoped regime and
disappears in overdoped samples, while the high-energy PG
persists in overdoped samples up to a hole doping
$p$$\approx$0.22. On the other hand, tunneling spectroscopy
data \cite{Renner1998}, whose PG energies correspond to the
high-energy scale in ARPES, showed that the PG exists in a
highly overdoped sample ($T_{c}$=56~K). Other measurements like
$c$-axis transport \cite{Shibauchi2001} also observed a PG for
an overdoped sample with $p$=0.22, though it is not clear
whether it is measuring the low-energy or high-energy PG.

Ultrafast time-domain pump-probe spectroscopy has shown to be a
useful tool in studying the nonequilibrium carrier dynamics in
HTSCs. This technique can differentiate between different
quasiparticle (QP) excitations by their different relaxation
timescales and thus distinguish the different phases, for
example, the PG phase in cuprate and pnictide HTSCs
\cite{Liu2008,Cao2008,Demsar1999,Chia2010}. In these
experiments, a pump pulse first breaks Cooper pairs into QPs
which rapidly relax to states close to the Fermi energy
($E_{F}$) by electron-electron and electron-phonon scattering.
The presence of a gap near $E_{F}$ causes a relaxation
bottleneck, so that carriers accumulate in states near the gap
edge, and subsequent relaxation and recombination dynamics give
rise to a transient change in optical transmission or
reflection of a time-delayed probe pulse which can be measured.
These studies helped in gaining information on the nature of
low-energy electronic structure of correlated electron systems
like HTSCs where the dynamics is sensitive to presence of a
gap.

Bi$_{2}$Sr$_{2}$CaCu$_{2}$O$_{8+\delta}$ (Bi-2212) is one of
the most studied HTSCs because of its extremely large
anisotropy and cleavability, containing only CuO$_{2}$ planes
and not chains, and the possibility of growing samples with a
larger range of $T_{c}$'s. The interpretation of the
femtosecond QP dynamics in Bi-2212 with regards to the PG phase
have only been reported for underdoped \cite{Liu2008} and
optimally doped \cite{Cao2008} samples. Liu \textit{et al.}
\cite{Liu2008} reported the coexistence of the PG and SC QPs in
the SC state of an underdoped sample by simultaneously
detecting two distinct components via tuning the probe beam
polarization and energy. Cao Ning \textit{et al.}
\cite{Cao2008} interpreted the dynamics of an optimally doped
sample using the Rothwarf-Taylor (RT) model and showed the
coexistence of a BCS-like temperature ($T$)-dependent SC gap
and a $T$-independent PG. In this paper, we analyze pump-probe
data of an overdoped (OD) Bi-2212 single crystal sample. In
fitting the $T$-dependence of the relaxation amplitude in the
SC state, we require the presence of \textit{two} $T$-dependent
gaps to fit the data. Our data are consistent with the scenario
where the SC gap and the PG coexist below $T_{c}$, i.e. the
``two-gap" scenario.

The experiments were performed on single crystals of Bi-2212
grown using the traveling-solvent-floating-zone method
\cite{Ichikawa1999}. The OD sample has been doped with Pb to
obtain a $T_{c}$ of 65~K ($p$=0.22). The value of $T_{c}$
obtained from magnetization data collected using Magnetic
Property Measurement System (MPMS) correspond to the midpoint
of the SC transition. The onset of superconductivity occurs at
68~K. The hole-doping value ($p$) were obtained from the
$T_{c}$ values using the parabolic law \cite{Presland91}
$T_{c}=T_{c}^{max}[1-82.6(p-0.16)^2]$, where
$T_{c}^{max}$=95~K. The sample was cleaved before data were
taken.

In our experiment, an 80-MHz Ti:Sapphire laser produces
sub-50~fs pulses at $\approx$ 800~nm (1.55~eV) as a source of
both pump and probe pulses. The pump and probe pulses were
cross-polarized. The pump spot diameter was 60~$\mu$m and that
of probe was 30~$\mu$m. The reflected probe beam was focused
onto an avalanche photodiode detector. The photoinduced change
in reflectivity ($\Delta R/R$) was measured using lock-in
detection. In order to minimize noise, the pump beam was
modulated at 100~kHz with an acousto-optical modulator. The
experiments were performed with an average pump power of
500~$\mu$W, giving a pump fluence of
$\sim$0.3~${\mu}$J/cm$^{2}$ and a photoexcited QP density of
$\sim1\times10^{-3}$/unit cell, showing that the system is in
the weak perturbation limit. The probe intensity was
approximately 10 times lower. The $T$ rise of the illuminated
spot has been accounted for in all the data.
\begin{figure}
\begin{center}
\includegraphics[width=16cm,clip]{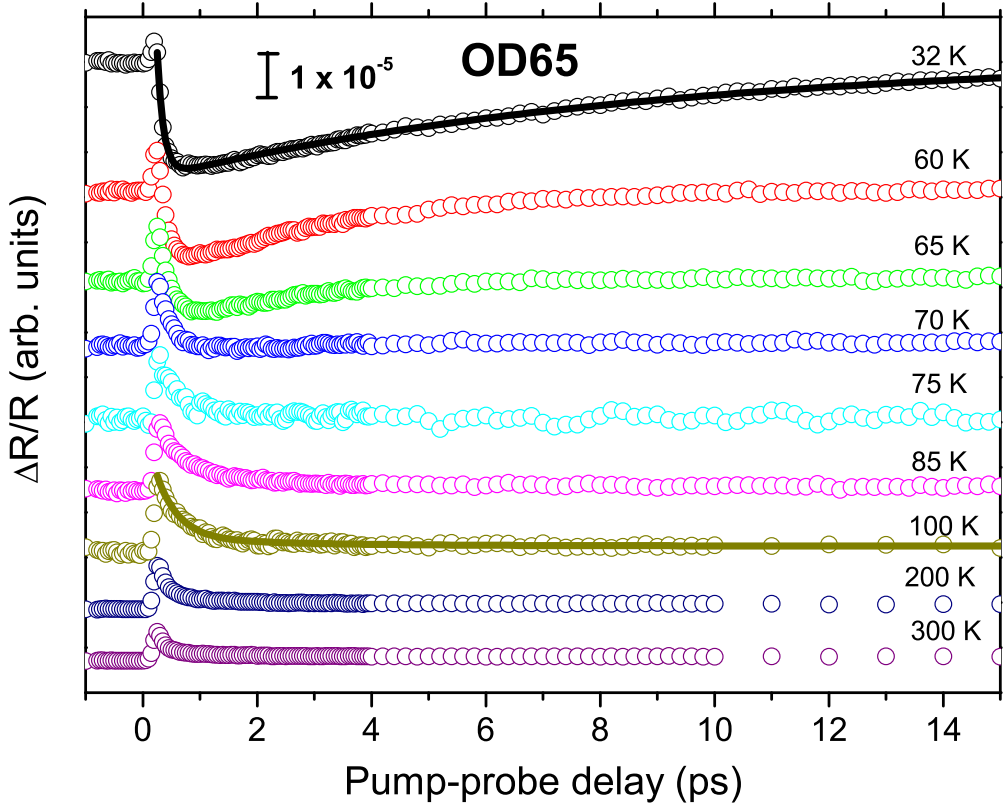}
\caption{(color online) Waterfall plots of transient reflection $\Delta R/R$
versus pump-probe time delay at different temperatures. Solid lines at 32~K and 100~K are
two-exponential fits.}
\label{fig:AllT}
\end{center}
\end{figure}

In Figure~\ref{fig:AllT} we show the time dependence of $\Delta
R/R$ at various temperatures above and below $T_{c}$. At low
temperatures, a fast \textit{positive} $\sim$100 femtosecond
(fs) decay ($A_{fast}$) and a slow picosecond (ps)
\textit{negative} decay ($A_{slow}$) were observed, with the
negative signal disappearing above $T_{c}$. We therefore
ascribe $A_{slow}$ to the reformation of SC order following
photoexcitation. Above $T_{c}$, a two-exponential positive
decay was seen up to 300~K.
\begin{figure}
\begin{center}
\includegraphics[width=12cm,clip]{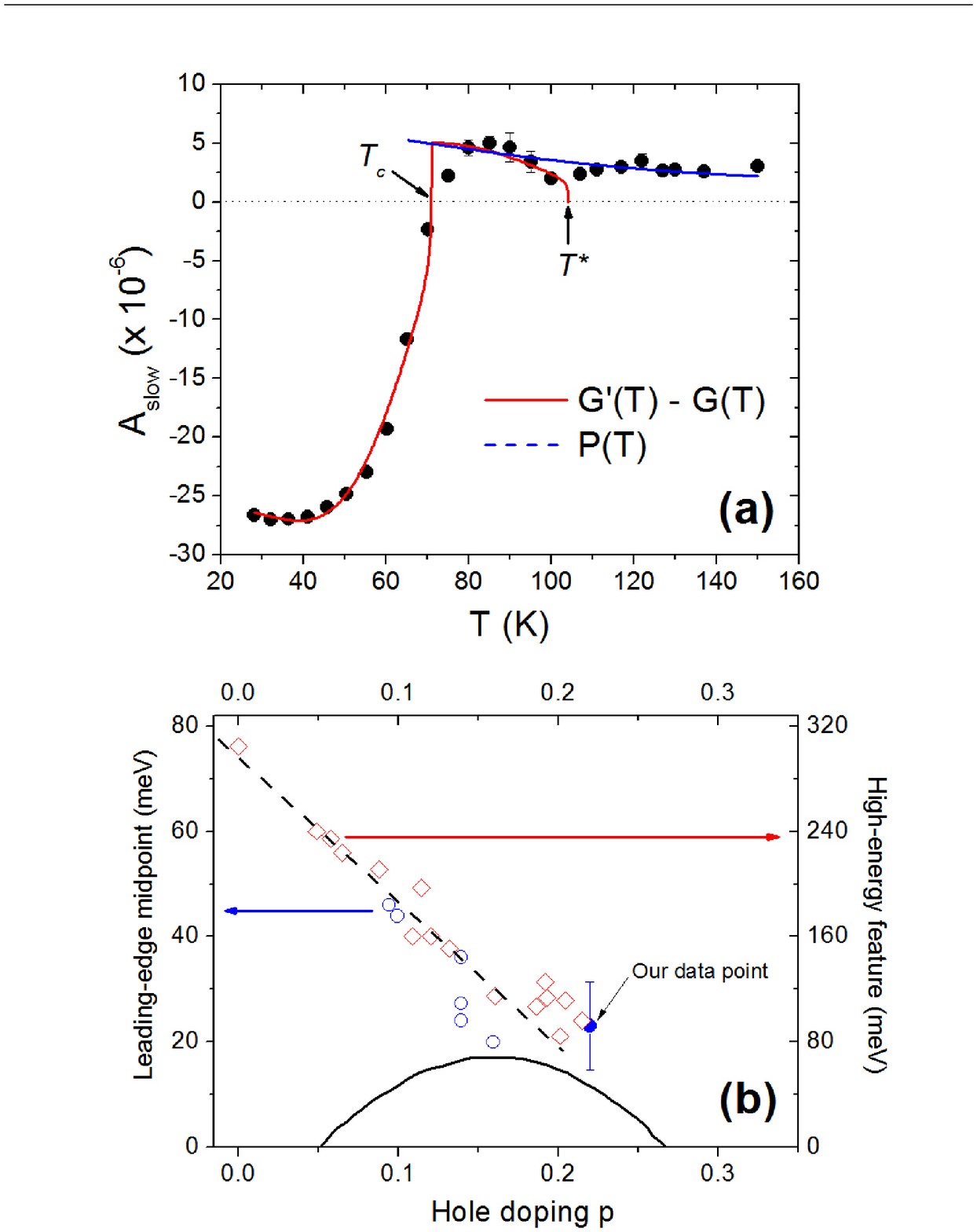}
\caption{(a) $T$ dependence of $A_{slow}$ (solid circles), with fit of data above
$T_{c}$ with $P(T)$ (dashed line), and fit of data from 28 K to $T^{\ast}$ with
$G^{\prime}(T) - G(T)$ (solid line). The fitted values of $T_{c}$ and $T^{\ast}$
are 71~K and 104~K, respectively. (b) Doping dependence of the PG as determined by the position of leading-edge midpoint ($\circ$, left axis) and high-energy feature ($\diamond$, right axis) in the ($\pi$,0) ARPES spectra from Bi-2212. The solid circle corresponds to the PG energy scale (22~meV) deduced from our data. The dome represents the $d$-wave mean-field approximation $\Delta (x) = 4.3 k_{B}T_{c}(x)/2$. Dashed line is a guide to the eye. Adapted with permission from Fig. 62 of Ref.~\onlinecite{Damascelli2003}. Copyright 2003 by the American Physical Society.} \label{fig:Amp}
\end{center}
\end{figure}

\begin{figure}
\begin{center}
\includegraphics[width=16cm,clip]{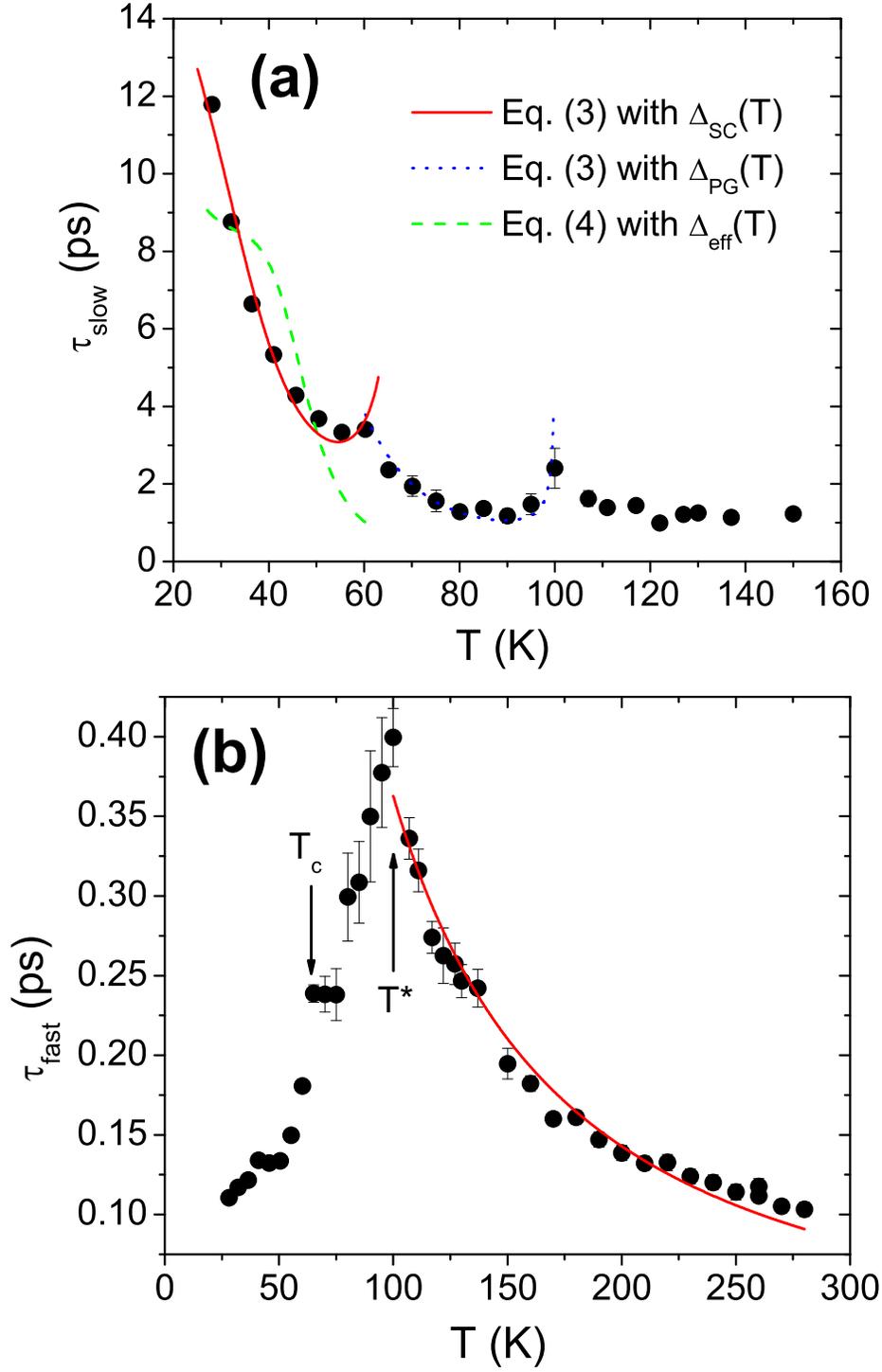}
\caption{(a) Solid circles: $\tau_{slow}(T)$ data. Solid line: Fit to
Eq.~(\ref{eqn:tau}) for $T<T_{c}$. Dotted line: Fit to
Eq.~(\ref{eqn:tau}) for $T_{c} < T < T^{\ast}$. Dashed line: Fit to
Eq.~(\ref{eqn:taueff}) for $T<T_{c}$. (b) $\tau_{fast}(T)$ data (solid circles), with fit to $1/T^{n}$
(solid line) where $n$=1.3.}
\label{fig:Tau}
\end{center}
\end{figure}

Using a two-exponential decay function, we extract the
temperature dependence of the relaxation amplitudes and
relaxation times, as shown in Fig.~\ref{fig:Amp}(a) and
Fig.~\ref{fig:Tau}. Notice that $A_{slow}$ (1) crosses zero at
$\sim$65~K, and (2) exhibits a dip at $\sim$100~K. The slow
relaxation time $\tau_{slow}$, on the other hand, exhibits an
upturn near two temperatures: (1) 65~K ($T_{c}$) and (2) 100~K,
which we denote as $T^{\ast}$, the PG temperature. Both the dip
in $A_{slow}$ at 100~K, and the upturn in $\tau_{slow}$ at 65~K
and 100~K, were reproducible upon re-cleaving the same sample.
In order to analyze the $T$-dependence of $A_{slow}$
quantitatively, we use the model proposed by Kabanov \textit{et
al.}~\cite{Kabanov1999}. The $T$-dependence of the relaxation
amplitude in the SC state for an isotropic
$T$-\textit{dependent} gap $\Delta_{c}(T)$ is given by
\begin{equation}
G(T)\propto\frac{\epsilon_{I}/(\Delta_{c}(T)+k_{B}T/2)}{1+\zeta
\sqrt{\frac{2k_{B}T}{\pi\Delta_{c}(T)}}\exp[-\Delta_{c}(T)/k_{B}T]}, \label{eqn:amplitudea}
\end{equation} where $\epsilon_{I}$ the pump laser intensity per unit cell, and $\zeta$
is a constant. The above expression for $G(T)$ describes a
reduction in the photoexcited QP density with increase in
temperature, due to the decrease in gap energy and
corresponding enhanced phonon emission during the initial
relaxation. On the other hand, the $T$-dependence of the
relaxation amplitude for a $T$-\textit{independent} gap
$\Delta_{p}$ is given by
\begin{equation}
P(T) \propto \frac{\epsilon_{I}/\Delta_{p}}{1 + \zeta
\exp(-\Delta_{p}/k_{B}T)}. \label{eqn:amplitudeb}
\end{equation} We first fit $A_{slow}(T>T_{c})$ with $P(T)$,
shown by the dashed line in Fig.~\ref{fig:Amp}(a) --- the fit
obviously does not reproduce the dip in $A_{slow}$ at 100~K.
Next, we proceeded to fit $A_{slow}(T)$ with the difference
$G^{\prime}(T) - G(T)$, where $G(T)$ is a function of the
$T$-dependent SC gap $\Delta_{SC}(T)$ which closes at $T_{c}$,
and $G^{\prime}(T)$ is a function of the $T$-\textit{dependent}
PG $\Delta_{PG}(T)$ which closes at $T^{\ast}$. Both
$\Delta_{SC}(T)$ and $\Delta_{PG}(T)$ are assumed to obey the
BCS $T$-dependence in this overdoped regime. The results are
shown as solid lines in Fig.~\ref{fig:Amp}(a). The dip in
$A_{slow}$ at 100~K is reproduced. The fitted values of $T_{c}$
and $T^{\ast}$ are 71~K and 104~K, respectively. We attribute
the discrepancy between the data and fitted lines near $T_{c}$
to fluctuation effects. Nevertheless the quality of the fits to
$\tau_{slow}(T)$ later are not affected. The fitted values of
the zero-temperature gaps are $\Delta_{SC}(0)=(3.0 \pm
0.2)k_{B}T_{c}$ and $\Delta_{PG}(0)=(4.1 \pm 1.5)k_{B}T_{c}$.
The fitted value of the zero-temperature SC gap
$\Delta_{SC}(0)$ agrees well with tunneling data
[$2\Delta_{SC}(0)/k_{B}T_{c}=5.3$] \cite{Hufner08}. Note that
our treatment of the PG to be $T$-dependent was motivated by
the dip of $A_{slow}$, and the concurrent upturn of
$\tau_{slow}$, at 100~K ($\approx$$T^{\ast}$). Compare this
with pump-probe data of other cuprates, such as
Y$_{1-x}$Ca$_{x}$Ba$_{2}$Cu$_{3}$O$_{7-\delta}$
\cite{Demsar1999}, HgBa$_{2}$Ca$_{2}$Cu$_{3}$O$_{8+\delta}$
\cite{Demsar2001}, and optimally doped Bi-2212 \cite{Cao2008},
where no dip in the relaxation amplitude, nor upturn in the
relaxation time, was seen at $T^{\ast}$, and so the PG's there
were treated to be $T$-independent. We attribute this
difference to the large amount of overdoping of our sample. In
tunneling data on an overdoped ($T_{c}$= 74.3~K) Bi-2212
sample, the PG has a smaller magnitude than the UD PG, and has
already almost vanished at 89~K \cite{Fischer2007,Renner1998}.
The attribution of a temperature dependence to our OD PG is
also consistent with the behavior of $\tau_{slow}$ near
$T^{\ast}$ --- the rapid appearance of a (pseudo)gap at
$T^{\ast}$ presents a relaxation bottleneck, which causes an
upturn in $\tau_{slow}$ at $T^{\ast}$. If the PG were to be
$T$-independent, there would not be a relaxation bottleneck,
and we would not have seen the upturn in $\tau_{slow}$ at
$T^{\ast}$. Our analysis suggests that, even in the OD regime,
there is a coexistence of the SC and PG phase below $T_{c}$.
Our results are thus consistent with the ``two-gap" scenario.
Our conclusions are also consistent with recent tunneling data
on an overdoped Bi-2201 sample \cite{Boyer2007}, where two gaps
were observed below $T_{c}$. Note that our value of
$\Delta_p(0)$$\approx$22~meV, is consistent with the trend in
the positions of the leading-edge mid-point of ARPES data at
lower dopings (see Fig.~\ref{fig:Amp}(b)). It is interesting to see
that our pump-probe technique yields values of the PG energy
scale that coincide with the low-energy PG scale from ARPES.

Next we analyze the $T$-dependence of the slow relaxation time
$\tau_{slow}(T)$ by using the RT model \cite{Rothwarf1967}.
This is a phenomenological model that describes the dynamics of
photoexcited QPs and high-frequency phonons (HFPs), where the
presence of a gap in the electronic density of states (DOS)
gives rise to a bottleneck for carrier relaxation. When two QPs
with energies $\geq$$\Delta$ ($\Delta$ is the gap magnitude)
recombine, a HFP is created with $\omega$$>$2$\Delta$. These
HFPs trapped within the excited volume can further rebreak
Cooper pairs and act as a bottleneck for QP recombination.
Hence the SC recovery is governed by the decay of the HFP
population. In the SC state ($T < T_{c}$), the $T$-dependence
of $\tau_{slow}^{-1}$ is determined by the $T$-dependence of
the amplitude $A_{slow}(T)$ and is given by
\cite{Kabanov2005,Cao2008,Chia2006}:
\begin{equation}
\tau_{slow}^{-1}(T)=\Gamma \left\{{\delta G(T)+\eta \sqrt{\Delta_{SC}(T)T}
\exp[-\Delta_{SC}(T)/T]}\right\}
\times
\left[\Delta_{SC}(T) + \alpha T \Delta_{SC}(T)^{4} \right],
\label{eqn:tau}
\end{equation} while in the PG phase ($T_{c} < T < T^{\ast}$),
$G(T)$ is replaced by $G^{\prime}(T)$, and $\Delta_{SC}(T)$
replaced by $\Delta_{PG}(T)$, with $\Gamma$, $\delta$, $\eta$
and $\alpha$ as fitting parameters. Figure~\ref{fig:Tau}(a)
shows the $T$-dependence of $\tau_{slow}$, and the fits given
by Eq.~(\ref{eqn:tau}) that reproduce the upturn of
$\tau_{slow}$ at $T_{c}$ and $T^{\ast}$. The good fits show
that the relaxation dynamics in both the SC phase and the PG
phase can be explained by the presence of a relaxation
bottleneck due to a gap in the DOS. The term
$[\Delta_{SC}(T)+{\alpha}T{\Delta_{SC}(T)}^4]$ in
Eq.~(\ref{eqn:tau}) accounts for the $T$-dependence of phonon
decay rate and ensures that the values of $\Delta_{SC}(0)$ and
$\Delta_{PG}(0)$ obtained from fits to $A_{slow}(T)$ and
$\tau_{slow}(T)$ are the same \cite{Chia2006}. Also note that
$\Delta_{PG}(0)/k_{B}T^{\ast}$=2.6
--- this ratio is consistent with the value (2.4) obtained from tunneling data
of an OD sample with $T_{c}$=82~K
\cite{Fischer2007,Dipasupil2002}, providing additional
justification that the crossover to the PG phase really does
take place at 100~K.

One might question the wisdom of using just $G(T)$ and
$\Delta_{SC}(T)$ component in Eq.~(\ref{eqn:tau}). After all,
in OD Bi-2212, the gap distribution on the sample surface is
more homogeneous than in underdoped or optimally-doped samples
\cite{McElroy05,Alldredge08}. One may suspect therefore that,
in the SC state, the SC gap and PG add in quadrature to yield
an effective gap $\Delta_{eff}(T) = \sqrt{\Delta^2_{SC}(T) +
\Delta^2_{PG}(T)}$ \cite{Chien09}. We attempt to fit the data
of $\tau_{slow}$ below $T_{c}$ using
\begin{equation}
\tau_{slow}^{-1}(T)=\Gamma \left\{{\delta [G(T) + G^{\prime}(T)]
+ \eta \sqrt{\Delta_{eff}(T)T}
\exp[-\Delta_{eff}(T)/T]}\right\}
\times
\left[{\Delta_{eff}(T)+{\alpha}T{\Delta_{eff}(T)}^{4}}\right].
\label{eqn:taueff}
\end{equation} The poor fit of Eq.~(\ref{eqn:taueff}) (dashed line) to data,
shown in Fig.~\ref{fig:Tau}(a), shows that this
``effective-gap" picture does not work. ARPES data, near the
antinodes of an OD Bi-2212 sample ($T_{c}$=86~K) \cite{Lee07},
are also inconsistent with the SC gap and PG adding in
quadrature.

We now turn to the fast component, which is positive at all
temperatures. A two-exponential positive decay was also seen in
optimally-doped Bi-2212 \cite{Perfetti2007} --- there the
authors attributed the fast decay to coupling between electrons
and ``hot" phonons (i.e. phonons that are strongly coupled to
the electrons), while the slow decay was due to anharmonic
coupling between the hot phonons and the cold lattice bath.
Figure~\ref{fig:Tau}(b) shows the temperature dependence of
$\tau_{fast}$ --- notice its rise with decreasing temperature,
before peaking at $T^{\ast}$ and decreasing to $\sim$100~fs at
30~K. Notice also the slight change in slope of $\tau_{fast}$
at $T_{c}$. The change in behavior of $\tau_{fast}$ at
$T^{\ast}$, and to a lesser extent at $T_{c}$, is intriguing
--- they suggest that the fast relaxation may result from an
admixture of electron-phonon \textit{and} electron-spin
fluctuation coupling. The peak at $T^{\ast}$, and its
subsequent decrease below $T^{\ast}$, may be due to an
increased scattering rate between electrons and spin
fluctuations as the sample enters the PG phase. This scenario
is further confirmed by the $T$-dependence of $\tau_{fast}$
above $T^{\ast}$ --- a fit to $1/T^{n}$ yields $n$=1.3, which
disagrees with the behavior predicted by Kabanov and Alexandrov
\cite{Kabanov2008} for the electron-phonon relaxation time for
good ($n$=2) and poor ($n$=3) metals.

We have performed ultrafast time-resolved photoinduced
reflectivity measurements on overdoped
Bi$_{2}$Sr$_{2}$CaCu$_{2}$O$_{8+\delta}$ single crystals. Our
data are consistent with the formation of a pseudogap phase at
$T^{\ast}$=100~K, which coexists with the superconducting phase
below $T_{c}$. We also see an increased scattering rate between
electrons and spin fluctuations as the sample enters the
pseudogap phase. Experimental studies on other
moderate-to-extreme overdoped cuprates are clearly needed to
confirm whether the pseudogap exists, is also
temperature-dependent, and whether $\tau_{fast}$ also peaks at
$T^{\ast}$, in these materials.

We acknowledge useful discussions with J. Demsar. This work was
carried out under the auspices of the NNSA of the U.S. DOE at
LANL under Contract No. DE-AC52-06NA25396, the LANL LDRD
program, EURYI, MEXT-CT-2006-039047, the Singapore Ministry of
Education Academic Research Fund Tier 1 (RG41/07) and Tier 2
(ARC23/08), and the National Research Foundation of Singapore.
This paper is written in memory of the late Professor Sung-Ik
Lee.

\bibliography{Bi2212OD70K}

\begin{thebibliography}{10}

\bibitem{Timusk1999}
T. Timusk and B. Statt, Rep. Prog. Phys. {\bf 62},  61  (1999).

\bibitem{Fischer2007}
{\O}. Fischer {\it et~al.}, Rev. Mod. Phys. {\bf 79},  353  (2007).

\bibitem{Panagopoulos98}
C. Panagopoulos and T. Xiang, Phys. Rev. Lett. {\bf 81},  2336  (1998).

\bibitem{Damascelli2003}
A. Damascelli, Z. Hussain, and Z.-X. Shen, Rev. Mod. Phys. {\bf 75},  473
  (2003).

\bibitem{Renner1998}
C. Renner {\it et~al.}, Phys. Rev. Lett. {\bf 80},  149  (1998).

\bibitem{Shibauchi2001}
T. Shibauchi {\it et~al.}, Phys. Rev. Lett. {\bf 86},  5763  (2001).

\bibitem{Liu2008}
Y.~H. Liu {\it et~al.}, Phys. Rev. Lett. {\bf 101},  137003  (2008).

\bibitem{Cao2008}
N. Cao {\it et~al.}, Chin. Phys. Lett. {\bf 25},  2257  (2008).

\bibitem{Demsar1999}
J. Demsar {\it et~al.}, Phys. Rev. Lett. {\bf 82},  4918  (1999).

\bibitem{Chia2010}
E.~E.~M. Chia {\it et~al.}, Phys. Rev. Lett. {\bf 104},  027003  (2010).

\bibitem{Ichikawa1999}
N. Ichikawa, Ph.D. thesis, University of Tokyo, 1999.

\bibitem{Presland91}
M.~R. Presland {\it et~al.}, Physica C {\bf 176},  95  (1991).

\bibitem{Kabanov1999}
V.~V. Kabanov, J. Demsar, B. Podobnik, and D. Mihailovic, Phys. Rev. B {\bf
  59},  1497  (1999).

\bibitem{Hufner08}
S. H{\"u}fner, M.~A. Hossain, A. Damascelli, and G.~A. Sawatzky, Rep. Prog.
  Phys. {\bf 71},  062501  (2008).

\bibitem{Demsar2001}
J. Demsar {\it et~al.}, Phys. Rev. B {\bf 63},  054519  (2001).

\bibitem{Boyer2007}
M.~C. Boyer {\it et~al.}, Nat. Phys. {\bf 3},  802  (2007).

\bibitem{Rothwarf1967}
A. Rothwarf and B. Taylor, Phys. Rev. Lett. {\bf 19},  27  (1967).

\bibitem{Kabanov2005}
V.~V. Kabanov, J. Demsar, and D. Mihailovic, Phys. Rev. Lett. {\bf 95},  147002
   (2005).

\bibitem{Chia2006}
E.~E.~M. Chia {\it et~al.}, Phys. Rev. B {\bf 74},  140409(R)  (2006).

\bibitem{Dipasupil2002}
R. Dipasupil, M. Oda, N. Momono, and M. Ido, J. Phys. Soc. Jpn. {\bf 71},  1535
   (2002).

\bibitem{McElroy05}
K. McElroy {\it et~al.}, Science {\bf 309},  1048  (2005).

\bibitem{Alldredge08}
J.~W. Alldredge {\it et~al.}, Nat. Phys. {\bf 4},  319  (2008).

\bibitem{Chien09}
C.-C. Chien, Y. He, Q. Chen, and K. Levin, Phys. Rev. B {\bf 79},  214527
  (2009).

\bibitem{Lee07}
W.~S. Lee {\it et~al.}, Nature {\bf 450},  81  (2007).

\bibitem{Perfetti2007}
L. Perfetti {\it et~al.}, Phys Rev Lett {\bf 99},  197001  (2007).

\bibitem{Kabanov2008}
V.~V. Kabanov and A.~S. Alexandrov, Phys. Rev. B {\bf 78},  174514  (2008).

\end{thebibliography}
\end{document}